# Pair distribution function study of Ni$_2$MnGa magnetic shape memory alloy: Evidence for the precursor state of the premartensite phase

Anupam K. Singh, Sanjay Singh,[*] and Dhananjai Pandey

*School of Materials Science and Technology, Institute of Technology (Banaras Hindu University) Varanasi, Uttar Pradesh, 221005, India*

Precursor phenomena preceding the martensite phase transition play a critical role in understanding the important technological properties of shape memory and magnetic shape memory alloys (MSMAs). Since the premartensite phase of Ni$_2$MnGa MSMA, earlier considered as the precursor state of the martensite phase, has in recent years been shown to be a thermodynamically stable phase, Singh *et al.* [Nat. Commun. **8**, 1006 (2017)], there is a need to revisit the precursor effects in these materials. We present here evidence for the existence of a precursor state of the premartensite phase in Ni$_2$MnGa MSMA by atomic pair distribution function analysis of high-energy, high-flux, and high-$Q$ synchrotron x-ray powder diffraction data. It is shown that the local structure of the cubic austenite phase corresponds to the short-range ordered (SRO) precursor state of the 3$M$ premartensite phase at temperatures well above the actual premartensite phase transition temperature $T_{PM}$ and even above the ferromagnetic (FM) transition temperature $T_C$. The presence of such a SRO precursor state of the premartensite phase is shown to lead to significant volume strain, which scales quadratically with spontaneous magnetization. The experimentally observed first-order character of the paramagnetic-to-FM phase transition and the anomalous reduction in the value of the magnetization in the temperature range $T_{PM} < T < T_C$ are explained in terms of the coupling of the magnetoelastic strain with the FM order parameter and the higher magnetocrystalline anisotropy of the precursor state of the premartensite phase, respectively.

## I. INTRODUCTION

Martensite transition is a diffusionless displacive phase transition where atoms in the high-symmetry cubic austenite phase move in a cooperative manner to lower the symmetry and give rise to a martensite phase with specific microstructural features [1–4]. This transition forms the basis of the development of several technologically important ferrous and nonferrous alloys with applications ranging from ground, marine, and aerospace structures to biomedical implants, energy conversion devices, actuators, and smart structures to name just a few [1–3]. Martensite transitions are ubiquitous and have been observed in different classes of materials, including elemental metals (Co and group-IV metals like Ti and Zr) [5,6], ferrous and nonferrous alloys [1–3,7], superconducting A-15 alloys [8], ceramics [2,3], oxides (ZrO$_2$) [9], sulfides (ZnS) [10], organometallic compounds [11], amino acids [12,13], and block copolymer micelles [14]. Martensite transition was first discovered in steels where it is irreversible and athermal [1]. In contrast, the temperature- and/or stress-induced martensite transitions in shape memory alloys (SMAs) like Ni-Ti [1–3,15,16], Ni-Al [1,17,18], and Cu-based alloys (like Cu-Zn-Al and Cu-Al-Ni) [2,19], and magnetic shape memory alloys (MSMAs) like Ni-Mn-Ga [20–22] are reversible and may have both athermal [1,23–25] and isothermal [1,2,24–27] characteristics, depending on the alloy system. The martensite transition is at the heart of the large and reversible shape change in the SMAs, wherein different orientational variants of the martensite phase, formed by twinning to maintain the invariant habit plane, merge on application of stress, while the original austenite state is recovered by heating to the austenite phase region [1–3]. The MSMAs have advantage over conventional SMAs, as they provide much larger strains not only under the influence of temperature and stress but also on application of magnetic field, which gives faster actuation within the martensite phase itself, but the brittleness of the alloys is still an issue of intense research [20–22,28–31].

Even though the shape change in the SMAs on the application of temperature or stress (or magnetic field in MSMAs) is due to the martensite phase transition, the martensite phase in the important SMAs exhibiting large strains does not result directly from the high-temperature cubic austenite phase but is preceded by an intermediate premartensite phase [1,2,32–37]. As a result, the precursor phenomena in SMAs and MSMAs have received enormous attention from the point of view of phonon softening, crystal structure, phase stabilities, and physical properties [32–48]. The precursor premartensite state in nearly stoichiometric or off-stoichiometric Ni-Ti or Fe-doped Ni-Ti [1,2,15,32,49] SMAs is characterized by the presence of diffuse scattering around the incommensurate positions of the trigonal structure (3$R$) [15,33,50–54]. This state is labeled as the premartensite-I [33] state which, at lower temperatures, is followed by the long-range ordered (LRO) $R$ phase, with a trigonal structure in the space group $P\bar{3}$ [16,39], also labeled as premartensite-II [33] or $R$-martensite phase [15,55,56]. In the case of Ni-Mn-Ga-based MSMAs, the premartensite phase is labeled as incommensurate 3$M$ modulated structure [$Immm(00\gamma)s00$ super-space group] with Burger orientation relationship $(110)_A||(010)_{PM}$, $[110]_A||[100]_{PM}$, $[1\bar{1}0]_A||[001]_{PM}$, and $[001]_A||[010]_{PM}$ for the lattice deformation shear [1,37,57–60]. The austenite-to-premartensite phase transition has been linked with complete (as in Ni-Ti SMA [15,61,62]) or partial softening (as in Ni$_2$MnGa MSMA) of the $\frac{1}{3}$ (110) transverse acoustic (TA$_2$) mode with displacement along the $[1\bar{1}0]$ direction observed in inelastic neutron scattering studies [41,42,44,45,63]. In the case of Ni$_2$MnGa MSMA, there is also Fermi surface nesting coupled to the soft mode that leads to Peierls-like lattice distortions and opens a pseudogap at the new Brillouin zone boundary of the premartensite phase [64,65].

---

*ssingh.mst@iitbhu.ac.in

Recent years have witnessed a lot of interest in understanding the precursor state of premartensite phase in the off-stoichiometric SMAs and MSMAs in relation to the occurrence of a fascinating strain glass state [48,55,56,66–72] which bears close analogy with spin glasses [73], dipole glasses [74], and relaxors [75]. In this context, one of the most investigated systems is Ti$_{50}$Ni$_{50-x}$Fe$_x$, which for low concentrations of Fe shows austenite $B2$ phase to incommensurate premartensite-I phase ($3R$) to LRO commensurate $R$-phase (premartensite-II) to LRO martensite $B19'$ phase (monoclinic) transitions with decreasing temperature [33,55,68,70]. Between the austenite and premartensite-II ($R$ phase) phase regions, the system shows a precursor state, comprising nanodomains of the $3R$ phase as heterophase fluctuations, whose stability regime widens with increasing Fe concentration until, for $x \geqslant 6$, the formation of the LRO $R$ and $B19'$ phases are completely suppressed and only the short-range ordered (SRO) premartensite phase survives [55,68,69,76]. The SRO premartensite phase exhibits locally fluctuating ferroelastic strains mimicking an ergodic strain liquid state whose characteristic time scale has been shown to diverge due to ergodic symmetry breaking at the strain glass freezing temperature ($T_g$) when the disorder content, and hence frustration, exceeds a critical value [66–68,76,77].

The stoichiometric Ni$_2$MnGa MSMA shows a close similarity with the sequence of thermodynamic phase transitions discussed above in the context of Ti$_{50}$Ni$_{50-x}$Fe$_x$ SMA for $x < 5$ [55,68]. With decreasing temperature, the $L2_1$ ordered austenite phase of Ni$_2$MnGa undergoes a first-order phase transition to LRO $3M$ modulated premartensite phase at $T_{PM} \sim 260$ K, which on further cooling undergoes a strong first-order phase transition to a $7M$ modulated martensite phase at $T_M \sim 210$ K [37,57,78]. Recent synchrotron x-ray powder diffraction (SXRPD) studies have confirmed the thermodynamic stability of the premartensite phase, both in stoichiometric and 10% Pt-substituted compositions, unambiguously [37,47]. However, the signature of the precursor phenomena in the parent cubic austenite phase in the form of phonon softening in inelastic neutron scattering [41] and diffuse scattering in elastic neutron scattering [42] studies well above the $T_{PM}$ necessitates a careful investigation of the local structure of Ni$_2$MnGa MSMA to understand the precursor state of the premartensite phase and its consequences on the magnetic phase transition in the cubic austenite phase.

In analogy with the SRO precursor state ($3R$) of the premartensite-II phase ($R$ phase) in the Ti$_{50}$Ni$_{50-x}$Fe$_x$ SMA [55,68], we present here evidence for the SRO precursor state of the premartensite phase in the cubic austenite phase of Ni$_2$MnGa with local $3M$-like structure well above the $T_{PM}$ and even above the paramagnetic-to-ferromagnetic (FM) transition temperature ($T_C$) using atomic pair distribution function (PDF) analysis. We show that the presence of such a SRO precursor state of the premartensite phase produces strains which couple with the FM order parameter fluctuations around $T_C$ and renormalize the coefficient of the fourth-order term in Landau expansion, leading to first-order character of the paramagnetic-to-FM phase transition, as confirmed by the observation of characteristic thermal hysteresis in the paramagnetic-to-FM transition in heating and cooling cycles. We also show that the temperature dependence of the unit cell volume of the austenite phase deviates significantly below the FM $T_C$ with respect to that in the paramagnetic austenite phase and that the excess volume scales quadratically with the FM order parameter, as expected for the magnetovolume effect [79,80]. We argue that the presence of the SRO precursor premartensite state in the FM phase is responsible for the reduction in the magnetization immediately below $T_C$, due to the higher magnetocrystalline anisotropy [81] of the premartensite phase, causing significant deviation from the $M \sim (T - T_C)^{1/2}$ type order parameter behavior expected for a second-order paramagnetic-to-FM phase transition [82,83]. This SRO precursor state of the premartensite phase is analogous to the unfrozen strain glass state or the so-called strain liquid state of Ni$_{50+x}$Ti$_{50-x}$ [77] with one significant difference. In Ni$_{50+x}$Ti$_{50-x}$ [77], the nanoscale domains in the strain liquid state do not transform to the LRO state, while the SRO precursor state of the premartensite phase in the stochiometric Ni$_2$MnGa gradually grows on lowering the temperature and leads to the LRO premartensite phase below $T_{PM}$.

## II. METHODS

Details of sample preparation and characterization used in this paper are described elsewhere [37]. The powder samples obtained from grinding the polycrystalline ingot used in this paper were annealed at 773 K for 10 h to remove residual stresses introduced during grinding [84–87]. The direct current (DC) magnetization measurements were carried out using a physical properties measurement system (Quantum Design), and alternating current (AC) susceptibility measurements were performed using SQUID-VSM (Quantum Design, MPMS). The AC susceptibility data were collected during the warming cycle on zero field cooled (ZFCW) and field cooled (FCW) states and during field cooling (FC) in the temperature range 2–400 K at 10 Oe (amplitude) and 333 Hz (frequency) with temperature sweep rate of 2 K/min. In the ZFCW protocol, the sample was cooled from 400 K (well above its FM $T_C$) down to 2 K in the absence of a magnetic field, and after that, the AC susceptibility was measured from 2 to 400 K during warming. Similarly, the data were collected in the FC and FCW cycles as well. DC magnetization data at 100 Oe field were also obtained in the temperature range 2–400 K under the ZFCW protocol using sweep rate of 4 K/min and in the temperature range 300–396 K under FC and FCW using sweep rate of 1 K/min. Additionally, DC magnetization data also obtained at 7 Tesla field in ZFCW cycle. The temperature-dependent SXRPD measurements were carried out in the high-resolution as well as high-$Q$ modes using high-energy x rays with a wavelength ($\lambda$) of 0.20706 Å at P02.1 beamline at PETRA III DESY, Germany. For the SXRPD measurements, borosilicate capillaries were used as sample containers. The high-$Q$ measurements were performed with a maximum instrumental $Q$ value ($Q_{\text{maxinst}}$) of 21.3 Å$^{-1}$. The high-$Q$ data for empty borosilicate capillary were also recorded for the background subtraction required for the conversion of the raw diffraction data into total scattering structure function $S(Q)$.

The average LRO structure was refined by the Rietveld technique [88] using the high-resolution SXRPD patterns. The refinement was carried out using the FULLPROF package [89]. For the refinement of the structure of the cubic austenite phase in the $Fm\bar{3}m$ space group, all atoms were considered at special positions, i.e., Ni at 8$c$ (0.25 0.25 0.25) and (0.75 0.75 0.75), Mn at 4$a$ (0 0 0), and Ga at 4$b$ (0.5 0.5 0.5) Wykoff positions, respectively [57]. The atomic PDF technique, originally developed to investigate the local correlations (i.e., SRO structure) in the liquids and amorphous solids, has become a powerful tool for investigating the deviation from the average LRO structure at short-range (SR) length scales [90–92]. It is a total scattering method in which both the Bragg and diffuse scattering, due to LRO and SRO structures, respectively, contribute simultaneously to the total scattering structure function $S(Q)$ [90–92]. The $S(Q)$ in the temperature range of 260–400 K were obtained from the high-$Q$ ($Q_{\max} = 21$ Å$^{-1}$) SXRPD data after applying the standard normalization and background corrections to raw scattering data using the PDFGETX3 program [93]. The experimental reduced atomic PDF, i.e., $G(r)$, was obtained using the program PDFGETX3 by taking the Fourier transform of the reduced structure function $F(Q)$, where $F(Q) = Q[S(Q) - 1]$, in the $Q$ range of $Q_{\min}$ to $Q_{\max}$ using the following equation:

$$G(r) = 4\pi r[\rho(r) - \rho_0] = \frac{2}{\pi} \int_{Q_{\min}}^{Q_{\max}} F(Q) \sin Qr \, dQ. \quad (1)$$

Here, $Q$ is the magnitude of the scattering vector $\mathbf{Q} = \mathbf{k}_i - \mathbf{k}_f$, where $\mathbf{k}_i$ and $\mathbf{k}_f$ are incident and reflected wave vectors, respectively, $Q_{\max}$ and $Q_{\min}$ are the maximum and minimum cutoffs, respectively, for the magnitude of the scattering wave vector, $\rho(r)$ is the atomic number density, and $\rho_0$ is the average atomic number density [90–92]. A sixth-order polynomial was used for modeling the background while processing the raw data, which results in the value of $r_{\text{poly}}$ to 0.9 Å during the Fourier transformation of $F(Q)$. The $r_{\text{poly}}$ is the distance in real space up to which an unphysical signal could be present due to possible errors during background modeling by the program PDFGETX3 with the assumption that there are no higher frequency aberrations in the raw data [93]. The structure refinement of the experimental atomic PDF was carried in the direct space using the program PDFGUI (PDFFIT2) [94], which is a promoted version of the program PDFFIT [95].

Since the present PDF data were collected using high-energy (60 keV) synchrotron x rays in the transmission mode, the x rays could penetrate through the powder sample in the sample holder and provide information about the bulk behavior, unlike the transmission electron microscopy techniques used earlier in the context of precursor state studies on off-stoichiometric Ni-Ti and Ni-Al alloys, which are sensitive to only thin regions of the sample near the surface [34,40,52–54,96]. The use of the high-flux synchrotron x-ray data, on the other hand, gives an excellent signal-to-noise ratio, which is a prerequisite condition for obtaining reliable atomic PDF data. Finally, the high-$Q$ SXRPD value helps us in minimizing the Fourier truncation ripples during Fourier transformation for obtaining the atomic PDF from the recorded high-$Q$ data [90].

## III. RESULTS AND DISCUSSION

### A. Magnetization

The temperature dependence of DC magnetization [$M(T)$] of Ni$_2$MnGa measured under an applied magnetic field of 100 Oe following the ZFCW protocol is shown in Fig. 1(a). The abrupt rise in $M(T)$ at $T \sim 210$ K is due to the martensite-to-premartensite phase transition, and the small dip at $T \sim 260$ K is due to the premartensite-to-austenite phase transition [see inset of Fig. 1(a)]. Further, a sharp drop in $M(T)$ at $T \sim 371$ K is due to the FM austenite-to-paramagnetic austenite phase transition at the Curie transition temperature $T_C \sim 371$ K. For the determination of the characteristic transition temperatures more precisely, we also measured AC susceptibility under the ZFCW, FC, and FCW protocols. The temperature dependence of the real part of AC susceptibility [$\chi'(T)$], measured at a temperature sweeping rate of 2 K/min, is shown in Fig. 1(b). At the Curie transition temperature $T_C \sim 371$ K for the FC cycle, there is a sharp increase in the $\chi'(T)$. On lowering the temperature further, a small drop in the $\chi'(T)$ is observed around the premartensite start transition temperature PM$_s^c \cong T_{\text{PM}} \cong 260$ K followed by a larger drop in $\chi'(T)$ at the martensite start transition temperature $M_s^c = T_M \cong 220$ K due to the increase in the magnetocrystalline anisotropy in the FM premartensite and martensite phases with respect to the FM austenite and premartensite phases, respectively, like that seen in Fig. 1(a) [97,98]. The nature of the $\chi'(T)$ curve in the FCW and FC protocols are similar, except that the characteristic transition temperatures during cooling (FC) are lower than those for the heating (FCW) cycle, as expected for a first-order phase transition [37,82,99]. The characteristic transition temperatures, austenite start ($A_s$), austenite finish ($A_f$), martensite start ($M_s$), and martensite finish ($M_f$), are $A_s \simeq 210$ K, $A_f \simeq 230$ K, $M_s \sim 220$ K, and $M_f \simeq 190$ K, respectively. The difference between the temperatures ($A_s + A_f$)/2, obtained during heating for the FCW protocol, and ($M_s + M_f$)/2, obtained during cooling for the FC protocol, gives the thermal hysteresis of $\sim 15$ K due to the first-order nature of the martensite phase transition [1,37,78,82]. A similar difference, though with a smaller value of thermal hysteresis ($\sim 2.4$ K), in the characteristic premartensite phase transition temperatures reveals the first-order character of the FM austenite-to-FM premartensite phase transition also. All the characteristic transition temperatures, determined from the $M(T)$ and $\chi'(T)$ plots, are in good agreement with those reported in the literature [37,44,57,60,81,86,97,98,100–102].

There are two intriguing features observed in the $M(T)$ and $\chi'(T)$ plots. The first one is the thermal hysteresis in $\chi'(T)$ around $T_C$, which is shown on an expanded scale in the inset (ii) of Fig. 1(b). The second one is a gradual decrease in the value of $M(T)$ below $T_C$ with decreasing temperature, as shown on a magnified scale in the inset of Fig. 1(a). The gradual drop in magnetization below $T_C$ has also been reported in the single crystalline samples of Ni$_2$MnGa [103]. The DC magnetization, which is the order parameter for the paramagnetic austenite to the FM austenite transition, should keep on increasing, albeit slowly as $(T - T_C)^{1/2}$ [82,83,104]. Moreover, since paramagnetic-to-FM phase transition is generally believed to be a second-order phase transition, one does not expect any thermal hysteresis around $T_C$ in Ni$_2$MnGa also. To further confirm the appearance of thermal hysteresis around $T_C$, we repeated the $M(T)$ measurements at 100 Oe DC field several times around $T_C$ with a slower temperature sweeping rate of 1 K/min. We give in Fig. 1(c) and in its inset the results of two such $M(T)$ measurements, labeled as I and II in Fig. 1(c), in the cooling and heating sequence. In all these measurements, a thermal hysteresis of $\sim 1.5$ K around $T_C$ [inset of Fig. 1(c)] was reproduced. This indicates that the paramagnetic-to-FM phase transition has a first-order character in Ni$_2$MnGa.

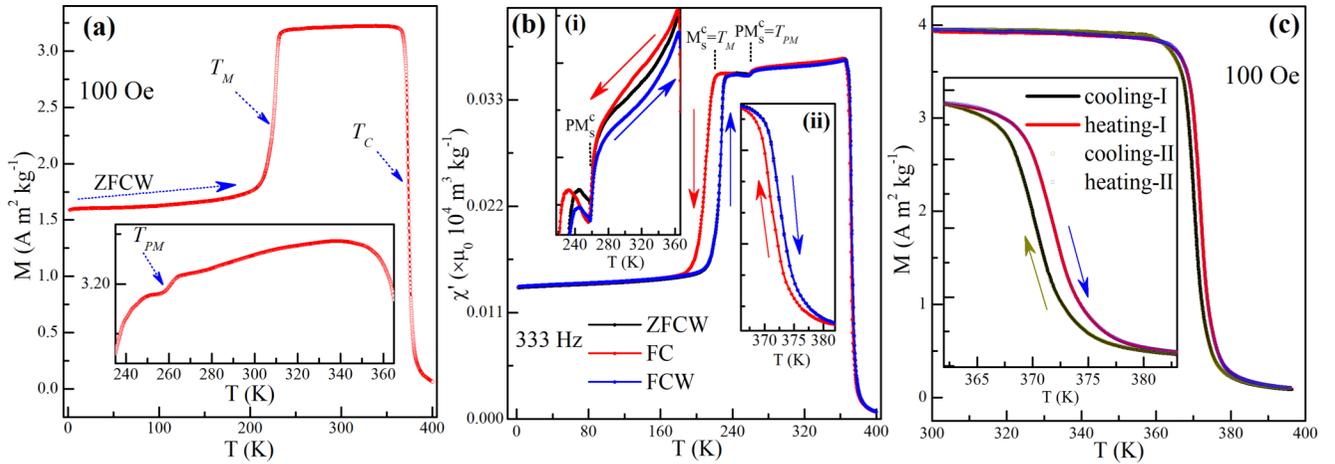

FIG. 1. Temperature dependence of the direct current (DC) magnetization of Ni$_2$MnGa, measured at 100 Oe in the zero field cooled warming (ZFCW) cycle at the rate of (a) 4 K/min and (c) 1 K/min and (b) temperature dependence of the real part of the alternating current (AC) susceptibility of Ni$_2$MnGa measured at 333 Hz in the ZFCW (black color), field cooled (FC; red color), and field cooled warming (FCW; blue color) cycles. The inset of (a) gives an enlarged view for 235 K $\leqslant T \leqslant$ 365 K range to clearly show the dip in magnetization at the premartensite phase transition temperature $T_{PM} \sim 260$ K as well as the anomalously decreasing behavior of magnetization below the ferromagnetic (FM) Curie temperature ($T_C$). The insets (i) and (ii) of (b) show an enlarged view around the premartensite and FM transition temperatures, respectively. The inset of (c) shows the thermal hysteresis across FM $T_C$ using two independent measurements labeled as I and II.

The observation of thermal hysteresis (first-order nature) across $T_C$ and the decrease in $M(T)$ at $T < T_C$ after its sharp jump around $T_C$ are not consistent with the conventional picture of a second-order paramagnetic-to-FM phase transition [82,83,105]. These observations encouraged us to perform detailed temperature-dependent structural studies above and below the $T_C$ for the average and local structures using high-resolution and high-$Q$ SXRPD data, respectively, to understand the genesis of the two unusual features in the magnetization studies.

### B. Temperature-dependent high-resolution SXRPD

The high-resolution SXRPD patterns in the temperature range 260–400 K, shown in Fig. 2(a), do not reveal any signature of a structural phase transition above the premartensite phase transition temperature $T_{PM} \sim 260$ K. The increasing shift in the position of the most intense Bragg peak toward the lower $2\theta$ side with increasing temperature, as shown in the inset (i) in Fig. 2(a), is apparently due to the usual thermal expansion behavior. The satellite peaks appearing around the most intense cubic Bragg peak at $T \leqslant 260$ K, shown more clearly in the inset (ii) of Fig. 2(a) at 260 K, correspond to the premartensite phase and are labeled as PM in the figure. It was verified by Rietveld refinement that all the Bragg peaks above $T_C$ as well in the temperature $T_{PM} < T \leqslant T_C$ in the SXRPD patterns are well accounted by the cubic austenite phase with the $Fm\bar{3}m$ space group. The Rietveld fits at 400 K $(> T_C)$ and at 270 K $(T_{PM} < T < T_C)$ are shown in Figs. 2(b) and 2(c), respectively. The refined lattice parameters are $a = 5.83064(3)$ Å and $5.81820(9)$ Å at 400 and 270 K, respectively, which are in good agreement with those reported in the literature [37]. The variation of the unit cell volume $(V)$, obtained from the Rietveld refinements, with temperature shows a linear dependence on temperature for $T < 370$ K and $T \geqslant 370$ K, as can be seen from Fig. 2(d). However, the slope of $V$ vs $T$ changes around the FM $T_C \sim 371$ K [Fig. 2(d)], obtained from the $M(T)$ and $\chi'(T)$ plots given in Figs. 1(a) and 1(b), respectively. The volume thermal expansion coefficient (TEC), defined as the ratio of change in volume ($\Delta V$), say from $V$ to $V + \Delta V$, on increasing the temperature ($\Delta T$), say from $T$ to $T + \Delta T$, to the change in temperature per unit the initial volume $V$ at $T$ : $\alpha_V(T) = \frac{\Delta V}{V * \Delta T}$ [106], was calculated from the temperature dependence of the unit cell volume, obtained by Rietveld refinement using the SXRPD data. It is interesting to note that the $\alpha_V$ exhibits a jumplike feature around $T_C$, as can be seen from the inset of Fig. 2(d). This feature is like that obtained using strain gauge measurements for the linear TEC on Ni$_2$MnGa [107].

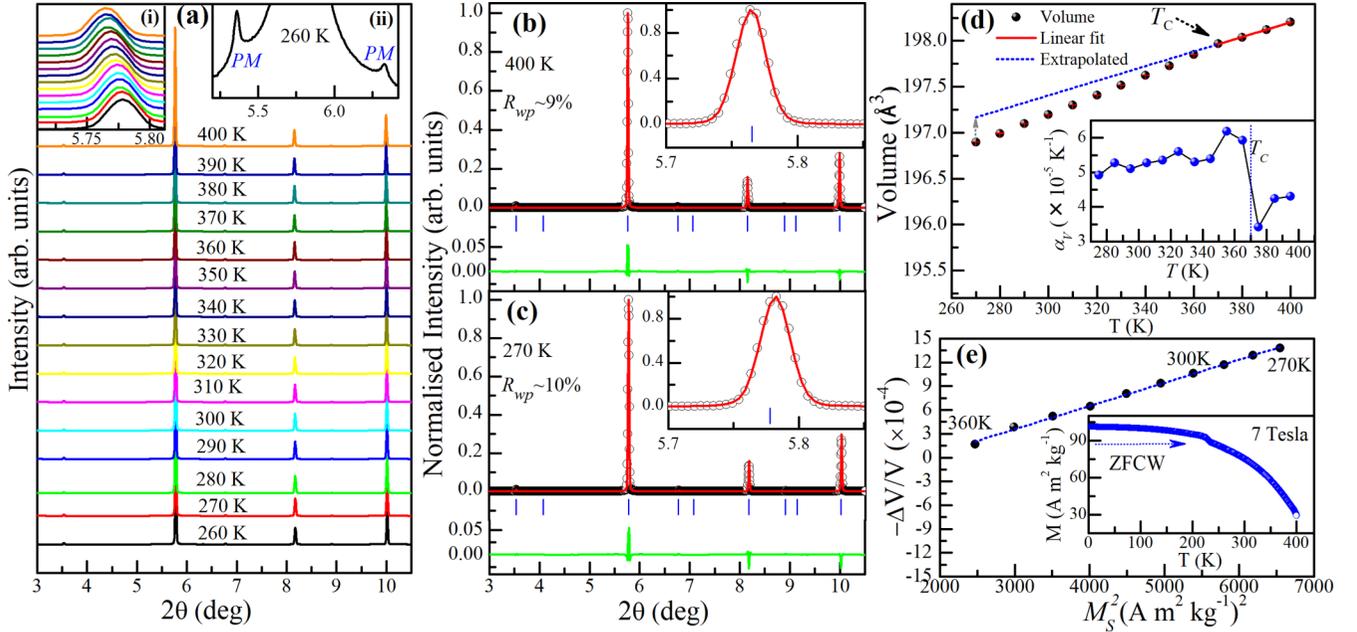

FIG. 2. (a) High-resolution synchrotron x-ray powder diffraction (SXRPD) patterns at various temperatures in the range 260–400 K. The inset (i) shows the enlarged view around the most intense Bragg peak of the cubic austenite phase at various temperatures, while the inset (ii) depicts a highly magnified view around the most intense peak at 260 K, showing the appearance of satellite peaks corresponding to the premartensite phase, which are marked as PM. (b) and (c) The observed (black circles), calculated (red continuous line), and difference profiles (green continuous line at the bottom) obtained after Rietveld refinement at 400 and 270 K, respectively, while the blue tick bars indicate the Bragg peak positions, and $R_{wp}$ is the weighted agreement factor. The insets in (b) and (c) show the quality of Rietveld fit around the most intense peak. (d) and (e) The variation of the unit cell volume ($V$) with temperature, showing volume contraction at $T \leqslant T_C$ and the variation of the cubic volume strain ($\Delta V/V$) with $M_s^2$ in the temperature range of 260 K $< T <$ 370 K, respectively. The dotted line (blue) in (d) is the extrapolated region of the linear expansion behavior above $T_C$, while the dotted line in (e) corresponds to the least squares linear fit. The insets of (d) and (e) depict the variation of the linear volume expansion coefficient ($\alpha_V$), obtained using the $V$ vs $T$ plot, and spontaneous magnetization $M_s$, measured at 7 Tesla field, as a function of temperature, respectively.

Since the Rietveld refinements using the SXRPD data reveal the absence of any structural phase transition around $T_C$, the change in the unit cell volume in the FM cubic austenite phase with respect to that in the paramagnetic cubic austenite phase suggests the presence of magnetoelastic coupling across $T_C$. The extrapolation of the linear volume expansion region of the paramagnetic cubic austenite phase, shown with blue line in Fig. 2(d), to the FM cubic austenite phase region (i.e., $T_{PM} < T < T_C$) reveals that the cubic unit cell volume undergoes substantial contraction below $T_C$ in the FM cubic austenite phase. Considering the presence of magnetoelastic terms, the Landau free energy functional for the FM transition can be written as [82,83,108]

$$\Delta G = \tfrac{1}{2}AM^2 + \tfrac{1}{4}BM^4 + \tfrac{1}{6}CM^6 + \tfrac{1}{2}kx^2 + QxM^2, \quad (2)$$

where $A = a_0(T - T_C)$, with $a_0$ being a temperature-independent constant. The first three terms in Eq. (2) correspond to the usual even powers of the order parameter ($M$) with temperature-dependent coefficients B and C. The fourth term ($\tfrac{1}{2}kx^2$) is the elastic energy term, where $k$ is the force constant and $x$ is the strain. The last term is the quadratic coupling term between strain and magnetization, with Q as the coupling coefficient. With $a_0$ and C as positive definite, the sign of the coefficient B decides the order of the phase transition [83,108]. On taking the first derivative of $\Delta G$ with regard to $x$ and considering a minimum stress condition in the material, i.e., $\frac{\partial \Delta G}{\partial x} = 0$, we obtain the following quadratic (magnetostrictive) relationship between the strain and the order parameter [83]:

$$x = -\frac{QM^2}{k}. \quad (3)$$

On eliminating the strain term ($x$) using the above relationship, Eq. (2) takes the form

$$\Delta G = \tfrac{1}{2}AM^2 + \tfrac{1}{4}\left(B - \frac{2Q^2}{k}\right)M^4 + \tfrac{1}{6}CM^6. \quad (4)$$

Evidently, the magnetoelastic coupling term ($\tfrac{1}{2}QxM^2$) in Eq. (2) renormalizes the coefficient of the $M^4$ term. If $\frac{2Q^2}{k} > B$, the renormalized coefficient of $M^4$ becomes negative, and Eq. (4) leads to a first-order phase transition [83,109]. Thus, the experimentally observed first-order character of the paramagnetic austenite-to-FM austenite phase transition is essentially due to the emergence of the magnetoelastic coupling in the Ni$_2$MnGa.

We calculated the change in the unit cell volume ($\Delta V$) at each temperature below $T_C$ in the FM region with respect to the extrapolated volume in the cubic paramagnetic region. The cubic volume strain ($\Delta V/V$) so obtained is found to scale quadratically with respect to the magnetization of the FM phase, as can be seen from Fig. 2(e), where the magnetization was obtained using high-field (7 Tesla) magnetization measurements as a function of temperature, shown in the inset of Fig. 2(e). At such a high field, one captures the spontaneous magnetization ($M_s$) of the Ni$_2$MnGa [110]. The quadratic dependence of $\Delta V/V$ on $M_s$ agrees with Eq. (3) that follows from the Landau free energy functional with the quadratic magnetoelastic coupling term in Eq. (2). This quadratic dependence of volume and longitudinal strain on magnetization has been termed as magnetovolume effect [79,80] and magnetostriction [111], respectively, in the literature.

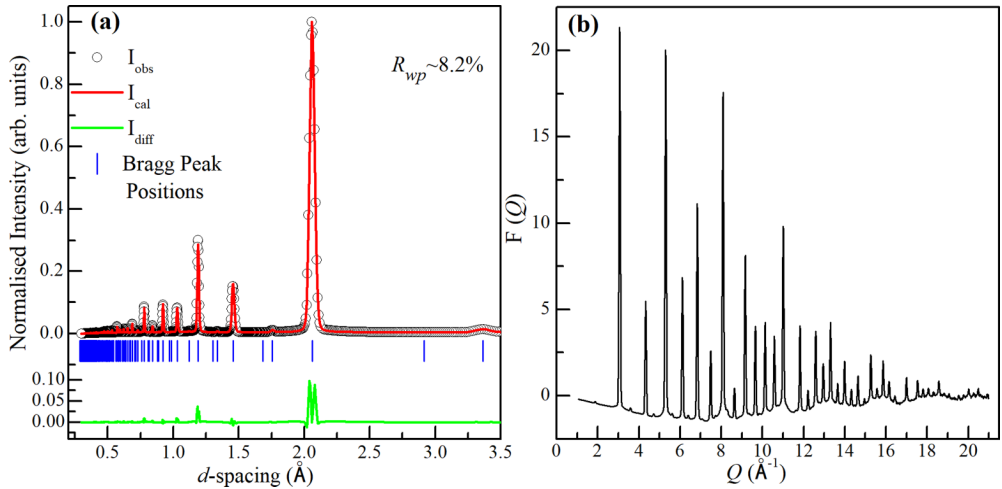

FIG. 3. (a) The observed (black circles), calculated (red continuous line) and difference profiles (green line), Bragg peak positions (blue ticks), and weighted agreement factor ($R_{wp}$) obtained after Rietveld refinement of the cubic austenite phase in the $Fm\bar{3}m$ space group using high-$Q$ synchrotron x-ray powder diffraction (SXRPD) data at 400 K. (b) The reduced structure function $F(Q)$ vs $Q$.

### C. Temperature-dependent atomic PDF analysis

Rietveld refinement captures the average LRO structure based on the Bragg peaks only and ignores the diffuse scattering. It cannot, therefore, capture the deviations from the LRO structure at short length scales of the order of one to a few unit cells which contribute to the diffuse scattering. An understanding of the emergence of such SR correlations can provide microscopic insight into the emergence of the magnetoelastic coupling in the cubic austenite phase. To explore both the SRO and LRO structures together, we decided to use the atomic PDF [90–93] technique using high-flux, high-energy, and high-$Q$ synchrotron x-ray data, the details of which will be presented in this section. The PDF has information not only about the SR regime (within one unit cell or so) but also the medium range (limited to a few unit cells) and long-range (LR, above the medium range in real space) regimes [112,113].

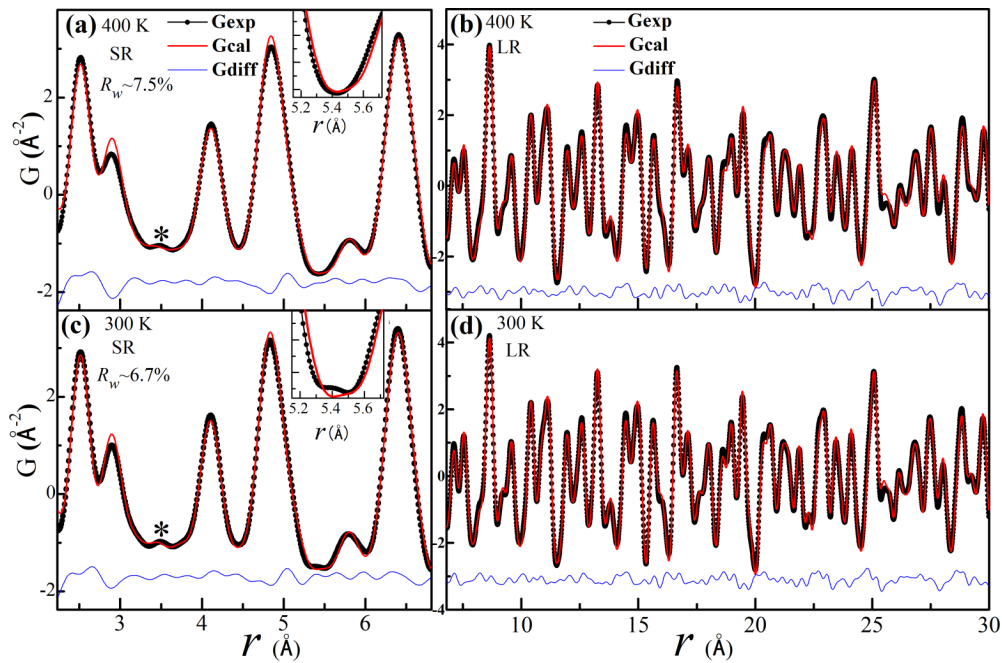

FIG. 4. The experimental (dark black dots connect with black line) and calculated (red continuous line) pair distribution functions (PDFs) and their difference (blue line at the bottom) obtained by real space structure refinement using cubic space group $Fm\bar{3}m$ in the (a) short-range (SR) and (b) long-range (LR) regimes at 400 K, and (c) SR and (d) LR regimes at 300 K. The insets in (a) and (c) show an enlarged view of the PDF fits at $r \sim 5.4$ Å. The asterisk (*) marked peak is a ripple due to the truncation of the Fourier series.

TABLE I. Interatomic distances corresponding to the refined structural parameters of austenite phase at 300 K and premartensite phase at 230 K reported in the literature [60].

| Austenite phase ($Fm\bar{3}m$) | | Premartensite phase ($Pnmn$) | |
|---|---|---|---|
| Atomic pair | Distance (Å) | Atomic pair | Distance (Å) |
| Ni-Mn and Ni-Ga | 2.52206 | Ni-Mn and Ni-Ga | 2.50–2.53 |
| Ni-Ni and Mn-Ga | 2.91222 | Ni-Ni and Mn-Ga | 2.84–2.97 |
| Ni-Ni, Mn-Mn, and Ga-Ga | 4.11851 | Ni-Ni, Mn-Mn, and Ga-Ga | 4.08–4.15 |
| Ni-Mn and Ni-Ga | 4.82938 | Ni-Mn and Ni-Ga | 4.75–4.90 |
| Ni-Ni and Mn-Ga | 5.04412 | Ni-Ni and Mn-Ga | 5.04–5.0422 |
| Ni-Ni, Mn-Mn, and Ga-Ga | 5.82445 | Ni-Ni, Mn-Mn, and Ga-Ga | 5.75–5.88 |
| Ni-Mn and Ni-Ga | 6.34705 | Ni-Mn and Ni-Ga | 6.28–6.39 |
| Ni-Ni and Mn-Ga | 6.51193 | Ni-Ni and Mn-Ga | 6.44–6.85 |

The results of the Rietveld refinement using high-$Q$ data at 400 K using the average cubic austenite structure are shown in Fig. 3(a), which shows an excellent fit between the observed and calculated peak profiles by accounting for all the Bragg peaks. Figure 3(b) shows the reduced structure function [$F(Q)$] with $Q_{max} = 21$ Å$^{-1}$ at 400 K. We note that the intensity of the peaks in the $F(Q)$ diminishes significantly toward higher $Q$ values, suggesting the dominance of the diffuse scattering. The corresponding reduced atomic PDF [i.e., $G(r)$] and its refinement at 400 K in the SR regime are shown in Fig. 4(a). There are several smaller intensity peaks present below the first PDF peak corresponding to the shortest Ni-Mn/Ga pair at $r \sim 2.52$ Å. These peaks appear during Fourier transformation of the raw PDF data due to the truncation of the Fourier series to a limited $Q$ value ($Q_{max} = 21$ Å$^{-1}$) and the background modeling [90,93,114]. These peaks do not correspond to any physical interatomic distances of the real space structure listed in Table I and are, therefore, considered as artifacts and hence not shown in Fig. 4 [90,93,114]. The small peak at $r \sim 3.5$ Å marked with an asterisk (*) in Fig. 4 also does not correspond to any physical pair of distances as per the crystallographic model listed in Table I. In fact, the wavelength of the peak at $r \sim 3.5$ Å matches with the wavelength of the termination ripple $\lambda = 2\pi/Q_{max}$ [90], with $Q_{max} = 21$ Å$^{-1}$. This confirms that the PDF peak at $r \sim 3.5$ Å is a noise/ripple. This peak gets fitted in Fig. 4 because the program PDFGUI considers the effect of Fourier truncation ripples also in the refinement [94].

The peak positions in the real space $G(r)$ (i.e., PDF) represent the pairs of atoms separated by a given distance $r$, and the peak width depends on the dynamic (thermal) and static disorder in the system [90,115]. The PDF peak shape provides information about atomic pair probability distribution, while the integrated intensity of the PDF peak is related to the coordination number [90–92]. All PDF peaks in the SR and LR regimes shown in Figs. 4(a) and 4(b), respectively, are well accounted for with the cubic structure in the $Fm\bar{3}m$ space group, confirming the average cubic austenite structure at high temperatures (400 K) at all length scales. However, we note the discrepancies between the calculated and observed PDF corresponding to second and fourth peaks centered at $r \sim 2.9$ and 4.9 Å, which correspond to the Ni-Ni, Mn-Ga, and Ni-Mn/Ga, Ni-Ni, Mn-Ga pairs of interatomic distances, respectively, covering various neighbors in the austenite phase (Table I). This misfit, as discussed later, is the first signature of local disorder in the SR regime even at a temperature as high as 400 K. The cubic lattice parameter obtained from the real space PDF refinement in LR regime turns out to be $a = 5.8258(4)$ Å at 400 K, which is in broad agreement with that obtained by the

TABLE II. Parameters obtained from the PDF refinement using cubic austenite ($Fm\bar{3}m$) and $3M$ commensurate premartensite ($Pnmn$) structures in the SR regime at selected temperatures. The $a$, $b$, and $c$ are the lattice parameters, $U_{iso}$ is the isotropic atomic displacements parameter, $\delta_2$ is the coefficient for $1/r^2$ contribution to the peak sharpening, and $R_w$ is the weighted agreement factor for the PDF refinement.

| | Austenite phase refined | | Premartensite phase refined | |
|---|---|---|---|---|
| Parameters | 400 K | 300 K | 260 K | 300 K |
| $a$ (Å) | 5.825 (6) | 5.811 (5) | 4.10 (1) | 4.11 (1) |
| $b$ (Å) | 5.825 (6) | 5.8117 (6) | 5.80 (1) | 5.805 (8) |
| $c$ (Å) | 5.825 (6) | 5.8117 (6) | 12.27 (3) | 12.28 (1) |
| $U_{iso}$ (Ni) | 0.013 (1) Å$^{-2}$ | 0.010 (1) Å$^{-2}$ | 0.002 (1) Å$^{-2}$ | 0.002 (1) Å$^{-2}$ |
| $U_{iso}$ (Mn) | 0.016 (3) Å$^{-2}$ | 0.016 (3) Å$^{-2}$ | 0.0003 (2) Å$^{-2}$ | 0.0004 (2) Å$^{-2}$ |
| $U_{iso}$ (Ga) | 0.007 (2) Å$^{-2}$ | 0.005 (1) Å$^{-2}$ | 0.004 (2) Å$^{-2}$ | 0.005 (3) Å$^{-2}$ |
| $\delta_2$ (Å$^2$) | 3.1 (0.4) | 3.2 (4) | 3 (1) | 5.60 (2) |
| $R_w$ (%) | 7.5 | 6.7 | 3.2 | 3.3 |

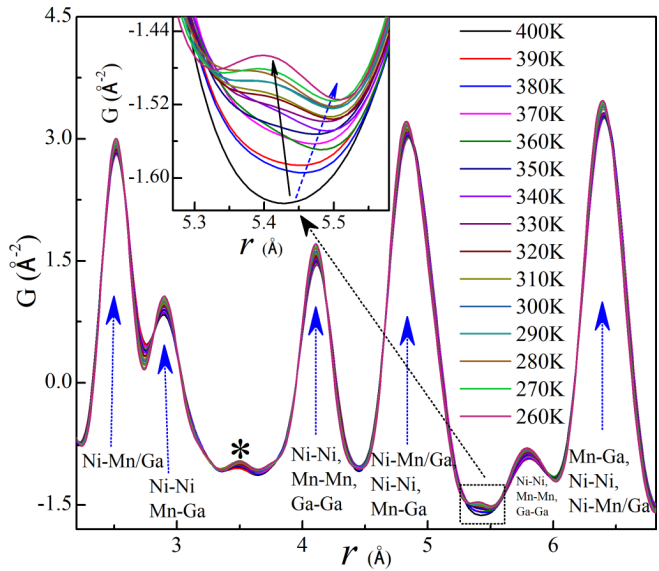

FIG. 5. Experimental pair distribution functions (PDFs) in the short-range (SR) regime at various temperatures from 400 to 260 K. The atomic pairs contributing to the individual peaks are indicated below the blue arrow line. The asterisk (*) marked peak is a ripple due to the truncation of the Fourier series. The inset shows an enlarged view at $r \sim 5.4$ Å, where the black arrow is a guide to the emergence of a new pair of interatomic distance, while the blue arrow shows the shift of the minima toward the higher $r$ side.

Rietveld refinement using the high-resolution SXRPD data at 400 K. After refining the structure in real space using $Fm\bar{3}m$ space group at 400 K, we attempted to refine the structure using the experimental $G(r)$ at 300 K assuming the same space group $Fm\bar{3}m$. Although the overall fit may appear satisfactory, it is evident from the inset of Fig. 4(c) that there is a misfit between calculated and experimental PDFs at $r \sim 5.4$ Å. Further, the misfit between the calculated and experimental PDFs around the second and fourth peaks persists, as at 400 K. On the other hand, the PDF fits in the LR region for $r \geqslant 6.8$ Å are quite good for the cubic structure, both at 400 and 300 K, as can be seen from Figs. 4(b) and 4(d). This reveals that the structure in the LR regime remains cubic at 400 and 300 K. However, the same structure gives rise to misfit between the calculated and the experimental PDF in the SR regime. The tiny peak at $r \sim 5.4$ Å, shown more clearly in the inset of Fig. 4(c) on a magnified scale, is obviously due to the departure from the cubic structure in the SR regime. The parameters obtained by refinement of the PDF in the SR regime, assuming the cubic structure, are listed in Table II at 400 and 300 K.

To understand the evolution of the local structure in the SR regime, we depict in Fig. 5 the atomic PDF in the temperature range 400–260 K covering the austenite and the premartensite phase regions. The atomic pairs corresponding to the various peaks in the PDF are marked with arrows in Fig. 5 based on the interatomic distances corresponding to the crystallographic model structure of the cubic austenite and 3$M$ premartensite phases listed in Table I [60]. Although there are no major changes observed in $G(r)$ in the temperature range 400–260 K (see Fig. 5), the temperature-dependent evolution of $G(r)$ at $r \sim 5.4$ Å, shown in the inset of Fig. 5, clearly reveals the emergence of a new peak. At 400 K, there is no such peak, but the shape of $G(r)$ begins to change on cooling $T < 400$ K, especially around the $T_C$ (~371 K). Further, for $T < 340$ K, the new PDF peak begins to appear. The change of shape of $G(r)$ followed by the emergence of this new peak at $r \sim 5.4$ Å on lowering the temperature suggests that this feature appears below $T_C$ ($\simeq 371$ K). Below $T_C$, the minimum in the PDF shifts from at $r \sim 5.45$ Å toward higher $r$ value, as shown schematically using the dotted blue arrow in the inset of Fig. 5. Further, the position of the peak at $r \sim 5.4$ Å shifts to the lower $r$ side with decreasing temperature, as shown by the solid black arrow in the inset of Fig. 5, and the intensity of this peak grows with decreasing temperature up to 260 K, which is the austenite-to-premartensite phase transition temperature $T_{PM}$ of Ni$_2$MnGa (see Fig. 1). Since the peak position in the atomic PDF represents the atomic pairs separated at a given distance $r$ for a given structure, the appearance of a new peak at $r \sim 5.4$ Å (see the inset of Fig. 5) with the change in temperature is a signature of the emergence of new atomic pairs. This indicates that the peak at $r \sim 5.4$ Å might be related to the local premartensite structure.

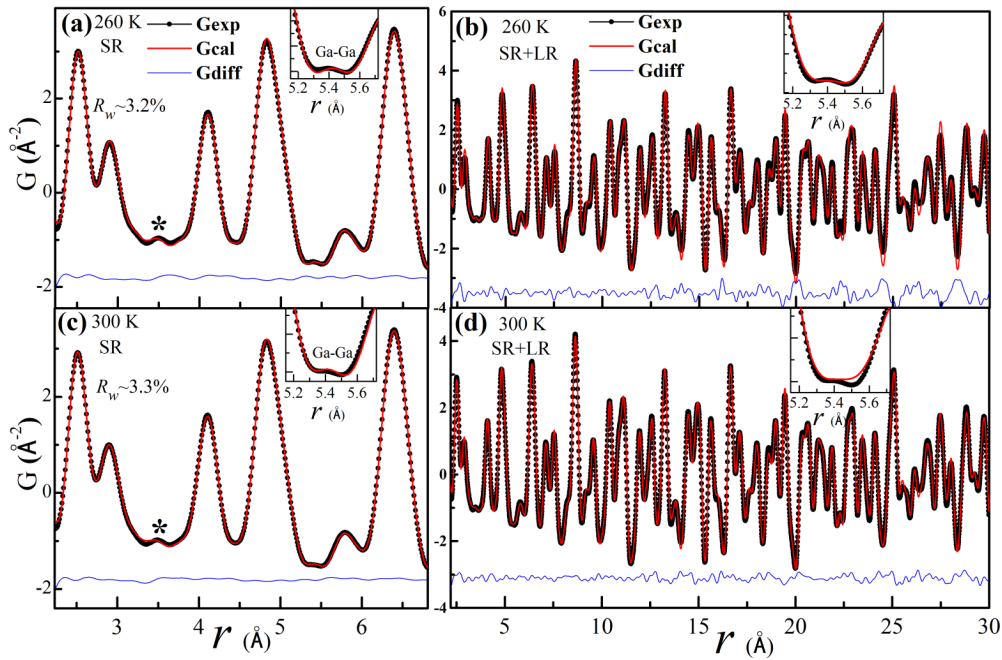

FIG. 6. The experimental (dark black dots connected with black line) and calculated (red continuous line) pair distribution functions (PDFs) and their difference (blue line at the bottom) obtained by real space structure refinement at 260 K using 3$M$ modulated orthorhombic premartensite phase space group $Pnmn$ in the (a) SR and (b) SR+LR regimes. The PDF fits at 300 K using the same space group ($Pnmn$) in the (c) SR and (d) the SR+LR regimes. The insets in (a)–(d) show an enlarged view of the PDF fit at $r \sim 5.4$ Å. The asterisk (*) marked peak is a ripple due to the truncation of the Fourier series.

To confirm the emergence of a premartensite structure, we first refined the structure using the experimental PDF data at 260 K, considering the $Pnmn$ space group (commensurate model) [60] of the premartensite phase. The results of the refinement are shown in Figs. 6(a) and 6(b) in the SR and SR + LR regimes, respectively. All PDF peaks are well accounted for, including the peak at $r \sim 5.4$ Å, which could not be accounted for at 300 K using cubic austenite structure [Fig. 4(c)]. Further, the misfit around the second and fourth PDF peaks (discussed in the context of Fig. 4) has completely disappeared. All these indicate that the misfits at $r \sim 2.9$ and 4.9 Å and the appearance of an extra peak at $r \sim 5.4$ Å (Figs. 4 and 5), which started appearing in the PDF well above $T_{PM}(\simeq 260$ K), are related to the premartensite phase. The real space PDF refined structural parameters for 260 K PDF data in the SR regime are given in Table II. After the PDF fitting, we found that the peak at $r \sim 5.4$ Å in PDF corresponds to the Ga-Ga atomic pair of the premartensite phase. The peak at $r \sim 2.9$ Å (second peak) corresponds to the Ni-Ni and Mn-Ga, while the peak at $r \sim 4.9$ Å (fourth peak) corresponds to the Ni-Mn/Ga, Ni-Ni, and Mn-Ga pairs of the premartensite phase. We investigated the emergence of features related to the premartensite phase in the SR regime as a function of temperature. Figures 6(c) and 6(d) depict the fits between the calculated and experimental PDF after refinement at 300 K using the structure of the premartensite phase in the $Pnmn$ space group in the SR and SR + LR regimes, respectively. The fits are excellent in the SR regime, as can be seen from Fig. 6(c). Interestingly, the extra peak at $r \sim 5.4$ Å is well captured with the premartensite structure model in the SR regime, as can be seen from the inset of Fig. 6(c). The refined parameters for the SR regime are listed in Table II. Attempts to fit both the SR and LR regimes together at 300 K led to misfit for the peak at $r \sim 5.4$ Å, as can be seen from the inset of Fig. 6(d). This clearly reveals that the signature of the premartensite phase is present only at SR length scales. In the LR regime, all peaks at 300 K were well fitted using the cubic $Fm\bar{3}m$ space group [see Fig. 4(d)]. Our results thus provide unambiguous evidence for the existence of the premartensite structure at 300 K in the SR regime, even though this temperature (300 K) is well above the actual premartensite transition temperature $T_{PM}(\simeq 260$ K). The feature in the experimental

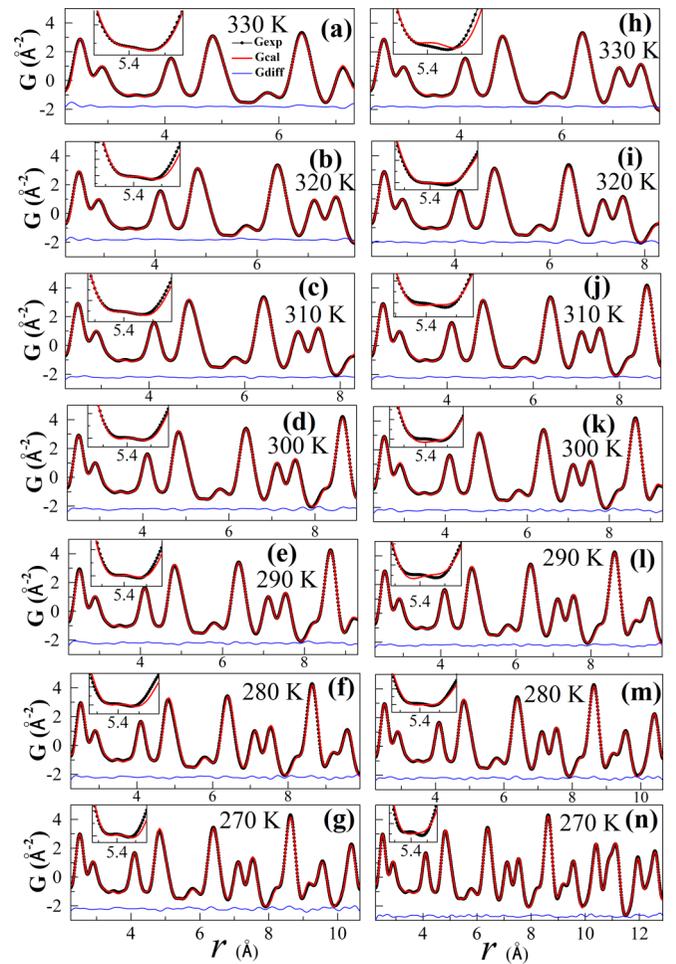

FIG. 7. The experimental (dark black dots connected with black line) and calculated (red continuous line) pair distribution functions (PDFs) and their difference (blue line at the bottom) using the 3$M$ modulated orthorhombic premartensite phase space group ($Pnmn$) in the temperature range 330–270 K obtained by real space structure refinement. The left panel [(a) to (g)] shows fits up to the correlation length ($\xi$) for which the premartensite phase structure can account for the peak at $r \sim 5.4$ Å, as shown in the insets. The right panel [(h) to (n)] depicts the fits to a distance which is greater than the correlation lengths. The misfit for the peak at $r \sim 5.4$ Å is quite evident from the insets of (h) to (n).

PDF that appears for $T < 370$ K (Fig. 5) itself confirms the precursor state of the premartensite phase within the austenite phase region. Our results are consistent with the observation of diffuse streaks due to such a precursor state in the austenite phase in the inelastic neutron scattering and high-resolution transmission electron microscopy (HRTEM) studies [41,42].

The length scale corresponding to the best fit to the experimental PDF is the correlation length ($\xi$) [82,113] of the precursor state of the premartensite phase. To determine the $\xi$ as a function of temperature, we gradually varied the range of $r$ up to which we could fit the experimental PDF using the premartensite phase structure. Figures 7(a)–7(g) correspond to the maximum distance in real space up to which premartensite phase structure can capture all PDF peaks successfully along with the anomalous peak at $r \sim 5.4$ Å. On the other hand, Figs. 7(h)–7(n) show that, when the value of $r$ exceeds the correlation length, a clear misfit at $r \sim 5.4$ Å is observed. The variation of $\xi$ with temperature is shown in Fig. 8, where the $\xi$ of the LRO premartensite phase below $T_{PM} \leqslant 260$ K was determined from the Scherrer equation [116] ($\xi = 0.9\lambda/\beta\cos\theta$) using the experimental full width at half maximum (FWHM; $\beta$) of the premartensite phase peak in the high-resolution SXRPD patterns around $2\theta = 6.32°$. The instrumental broadening was negligible as compared with the observed FWHM. The discontinuous behavior of $\xi$ at 260 K with its saturation for $T < 260$ K, seen in Fig. 8, clearly reveals the first-order nature of the austenite-to-premartensite transition temperature [43,98]. The presence of the tail region for $T > 260$ K, where $\xi$ grows very slowly as a precursor state of the premartensite phase with decreasing temperature, suggests that this transition is a fluctuation-driven first-order phase transition [117]. We find that $\xi$ increases from $r \sim 7.4$ Å in the austenite phase region at 330 K to $r \sim 367$ Å at 230 K in the premartenste phase region. It is interesting to note that the $\xi$ obtained by us above $T_{PM}$ ($\sim 1$ nm at 290 K) is comparable with the width of alternate dark and light bands ($\sim 1.2$ nm at 293 K) in HRTEM images of the tweed microstructure, which is the characteristic feature of the precursor state of the premartensite phase in Ni$_2$MnGa [41]. The analysis of experimental PDF thus reveals that the local structure of the cubic austenite phase corresponds to the precursor state of the premartensite phase and that this precursor state appears well above $T_{PM}$ as well as above the FM $T_C$. The presence of the precursor state of the premartensite phase in the SR regime of the austenite phase is responsible for the emergence of the magnetoelastic strains observed in the FM austenite phase region, which in turn leads to the first-order character of the paramagnetic-to-FM phase transition due to the coupling of the FM order parameter with the magnetoelastic strains [37,43]. Also, the local premartensite structure of the cubic austenite phase is responsible for the decrease in magnetization [Fig. 1(a)] below $T_C$ due to higher magnetocrystalline anisotropy energy of the premartensite phase than the cubic austenite phase [81].

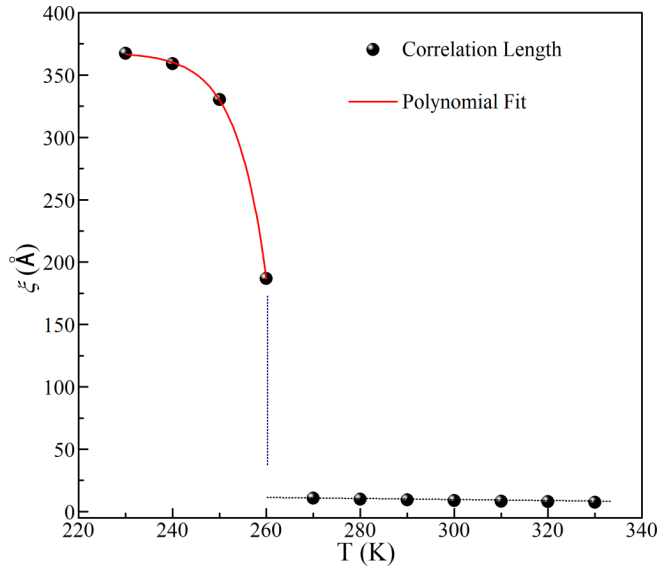

FIG. 8. The correlation length ($\xi$) of the premartensite phase as a function of temperature. The black dotted line shows the slowly increasing trend of $\xi$ in the temperature range 330–270 K. The sharp increase in $\xi$ at $T_{PM} \sim 260$ K is shown by the dotted blue line. Below 260 K, $\xi$ tends toward its saturation value for the long-range ordered premartensite phase.

## IV. CONCLUSIONS

To conclude, we have presented here the evidence for the existence of the precursor state of the premartensite phase in the Ni$_2$MnGa MSMA by atomic PDF analysis of the high-$Q$ SXRPD data. Our results reveal that the local structure of the cubic austenite phase corresponds to the precursor state of the premartensite phase. We have also presented evidence for quadratic magnetoelastic coupling in the austenite phase arising from the presence of such a SRO precursor state. Within the framework of the Landau theory, the coupling of the magnetoelastic strains with the FM order parameter around $T_C$ imparts first-order character to the paramagnetic-to-FM phase transition, as revealed by the thermal hysteresis in the FM $T_C$ in the heating and cooling magnetization cycles. The presence of the local premartensite precursor state in the FM austenite phase also explains the anomalous reduction in the magnetization below $T_C$ due to the higher magnetocrystalline anisotropy of the premartensite phase causing significant deviation from the $M \sim (T - T_C)^{1/2}$ type order parameter behavior expected for a second-order FM phase transition. The present findings significantly advance the understanding of the precursor state of the premartensite phase and its impact on the nature of paramagnetic-to-FM phase transition as well as the temperature dependence of the magnetization in the FM state of MSMAs in general and Ni$_2$MnGa in particular.

## ACKNOWLEDGMENTS

S.S. thanks C. Felser, R. Rawat, and S. R. Barman for useful discussions. S.S. is thankful to the Science and Engineering Research Board of India for financial support through the award of Ramanujan Fellowship (Grant No. SB/S2IRJN-015/2017) and Early Career Research Award (Grant No. ECR/2017/003186) and UGC-DAE CSR, Indore for financial support through its "CRS" Scheme. Portions of this research were conducted at the light source PETRA III of DESY, a member of the Helmholtz Association. Financial support from the Department of Science and Technology, Government of India within the framework of the India@DESY is gratefully acknowledged. We would like to thank the beamline scientist Dr. Martin Etter for his help in setting up the experiments.